\documentclass[superscriptaddress,twocolumn,noshowpacs,pre,amsmath,amssymb, amsfonts]{revtex4-1}   

\usepackage{graphicx}
\usepackage{dcolumn}
\usepackage{bm}
\usepackage{colordvi}
\usepackage{epsfig}
\usepackage{graphicx}
\usepackage[usenames]{color}
\usepackage{epstopdf}

\usepackage{hyperref} 
\hypersetup{
  colorlinks,
  citecolor=blue,
  linkcolor=blue,
  urlcolor=blue
  }
  
\usepackage[usenames,dvipsnames]{xcolor}
\usepackage[normalem]{ulem}

\usepackage{siunitx}
\usepackage{comment}

\usepackage{glossaries}
\newacronym{RFOT}{RFOT}{random first-order theory}
\newacronym{PTS}{PTS}{point-to-set}
\newacronym{MD}{MD}{molecular dynamics}
\newacronym{LJBM}{LJBM}{Lennard-Jones binary mixture}
\newacronym{EAM}{EAM}{embedded-atom method}
\newacronym{MCT}{MCT}{mode-coupling theory}
\newacronym{SBR}{SBR}{stochastic $\beta$-relaxation theory}
\newacronym{ISF}{ISF}{intermediate scattering function}
\newacronym{SISF}{SISF}{self-\acrlong{ISF}}
\newacronym{SE}{SE}{Stokes--Einstein}
\newacronym{CRR}{CRR}{cooperative rearrangement regions}

\begin{document}

\title{Non-monotonic dynamic correlations beneath the surface of glass-forming liquids}
\author{Hailong Peng}
\email[]{hailong.peng@csu.edu.cn}
\affiliation{School of Materials Science and Engineering, Central South University, 932 South Lushan Rd, 410083 Changsha, China}
\author{Huashan Liu}
\affiliation{School of Materials Science and Engineering, Central South University, 932 South Lushan Rd, 410083 Changsha, China}
\author{Thomas Voigtmann}
\email[]{thomas.voigtmann@dlr.de}
\affiliation{Institut f\"ur Materialphysik im Weltraum, Deutsches Zentrum f\"ur Luft- und Raumfahrt (DLR), 51170 K\"oln, Germany}
\affiliation{Department of Physics, Heinrich-Heine-Universit\"at D\"usseldorf, Universit\"atsstra\ss{}e 1, 40225 D\"usseldorf, Germany}

\begin{abstract}
Collective motion over increasing length scales is a signature of the vitrification process of liquids.
We demonstrate how distinct static and dynamic length scales
govern the dynamics of vitrifying films.
In contrast to a monotonically growing static correlation length, the dynamic correlation length that measures the extent of surface-dynamics acceleration into the bulk, displays a striking non-monotonic temperature evolution that is robust also against changes in detailed interatomic interaction. 
This non-monotonic change defines a cross-over temperature $T_*$ that is
distinct from the critical temperature $T_c$ of \acrfull{MCT}.
We connect this non-monotonic change to a cross-over from mean-field like
liquid dynamics to glass-like dynamics that is signalled by a
morphological change of \acrfull{CRR} of fast particles, and as the point
where fast-particle motion decouples from structural relaxation.
We propose a rigorous definition of this new cross-over temperature $T_*$
within a recent extension of \acrshort{MCT}, the \acrfull{SBR}.
\end{abstract}
\date{\today}
\maketitle

Dynamical processes
in a liquid close to the glass transition become cooperative across spatial regions of increasing extent \cite{Berthier:2011}, and it is thus natural
to seek an intrinsic correlation length whose divergence would signal
the transition.
But the hallmark of the glass transition is a dramatic change in the dynamics
that is caused by only weak changes in the statics.
Consistently, attempts at defining \emph{static} correlation lengths have
found only weak changes in these quantities close to the (computationally
or experimentally accesible part) of the transition \cite{Ediger:2000,Berthier:2011,Flenner:2014,Bennemann:1999}.
Only recently it has become clear that in certain perturbed systems,
\emph{dynamic} correlation lengths can be defined that
display a much more interesting, non-monotonic behavior \cite{Kob:2012,Hocky:2014,Nagamanasa:2015} with a peak at some cross-over temperature.
Such non-monotonic variations near the glass transition have since emerged
as a signature of various non-equilibrium glass-forming systems
\cite{Lozano:2019,Li:2020}.

The prevailing methodology to detect spatial correlations in glassy systems is suggested by the \gls{RFOT} \cite{Kirkpatrick:1989,Cavagna:2007,Biroli:2008}:
pinning a subset of particles in the equilibrium fluid, one examines how the configuration of the rest of particles is influenced
\cite{Scheidler:2003,Franz:2007,Zhang:2016,Balbuena:2019}.
While this \gls{PTS} protocol is designed to keep the \emph{static} properties of the system in equilibrium,
it represents a strong perturbation of the dynamics \cite{Flenner:2012}:
The freezing of some particles can be viewed as imposing a zero-temperature region and hence a strong temperature gradient, yet the associated linear-response regime shrinks to zero at the glass transition \cite{Vaibhav:2020}.
We propose the study of glassy films as a new method to detect such dynamic correlation lengths by measuring how far the statics and dynamics into the bulk liquid affected by the accelerated mobility on the surface.

We show here that the dynamics in \emph{equilibrated films} is similarly governed by a non-monotonically dynamical correlation
length.
This opens an interesting link between the study of such films and our
fundamental understanding of the glass transition in the bulk, in analogy to the situation that the diverging width of a gas--liquid interface at its critical point is governed by the same diverging correlation length in the bulk \cite{Jasnow:1984}.
Furthermore the free-surface dynamics is an important factor in the
preparation of glassy films
\cite{Fakhraai:2008,Zhu:2011,Tanis:2019}, and more specifically
ultra-stable glasses \cite{Swallen:2007,Yu:2013,Singh:2013,Berthier:2017,Luo:2018} and nanostructured materials \cite{Chen:2017} that are produced by depositing atoms layer-by-layer on an amorphous substrate.

We demonstrate that a specific cross-over temperature $T_*$ governs both the point of maximal dynamical correlations in the film geometry,
and the point where \gls{CRR} of fast particles in the bulk undergo a shape transition.
This cross-over point $T_*$ is significantly above the critical temperature $T_c$ of the \gls{MCT}. We rationalize this new cross-over point in the context of a recent extension of \gls{MCT}, the \gls{SBR}, as the point where
the effect of long-range fluctuations in the dynamical order parameter
of the theory is most pronounced in decoupling fast-particle dynamics from
bulk relaxation.

\begin{figure*}
\includegraphics[width=0.8\linewidth]{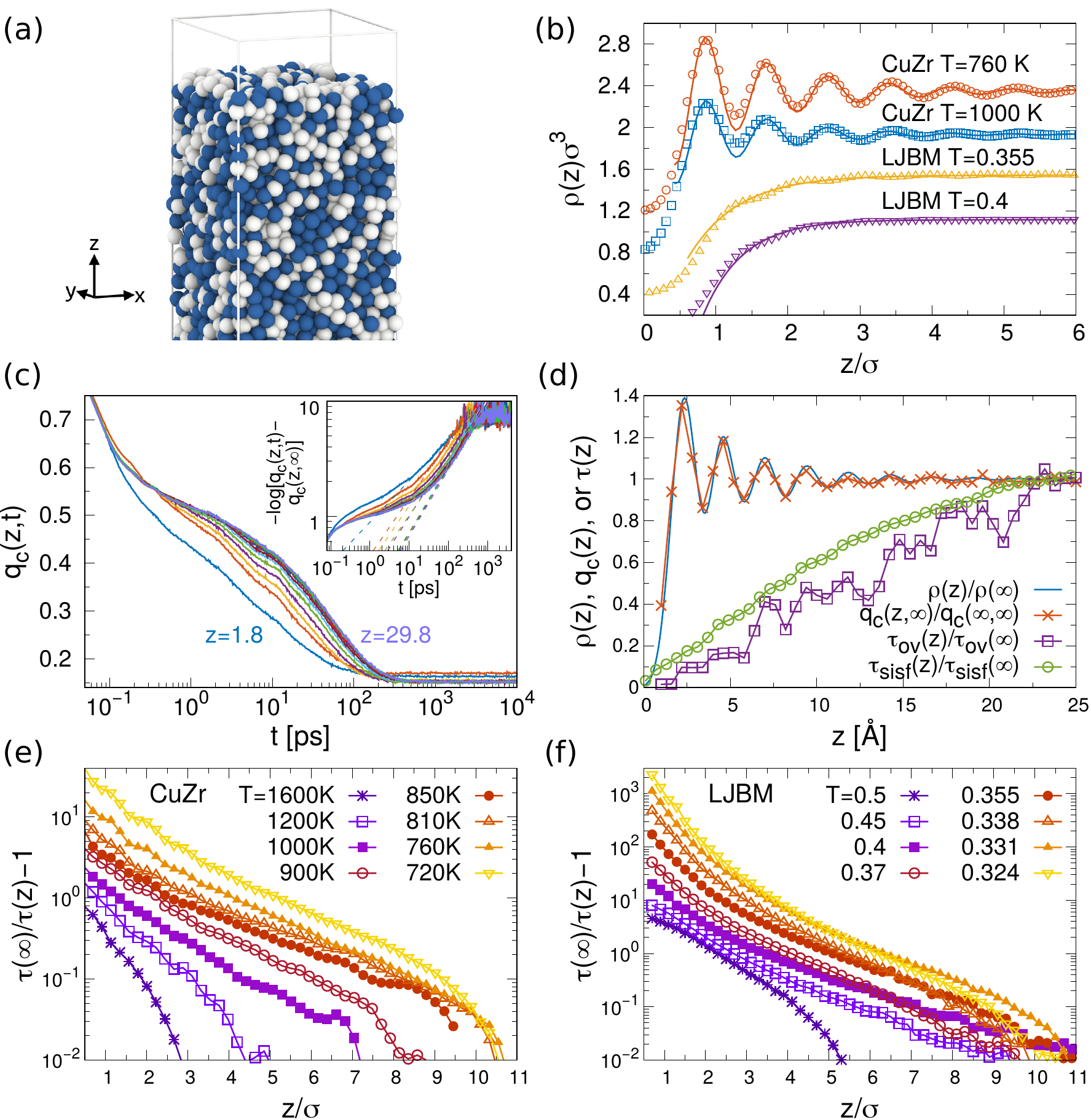}
\caption{Characterization of dynamic properties near the surface.
(a) Illustrative snapshot of the simulation setup (CuZr system, colors indicating atomic species).
(b) Static density profiles $\rho(z)$ as a function
  of distance $z$ from the surface along the normal into the bulk, for
  the CuZr liquid and the \acrfull{LJBM} (in units of the average atomic radius $\sigma$, shifted vertically in steps of $0.4/\sigma^3$ for clarity).
 Solid lines are fits to extract the static correlation length.
(c)
Representative decay of the overlap correlation function $q_c(z,t)$ with the distance $z$ to the surface (CuZr; $T=\SI{810}{\kelvin}$).
Dashed lines in the inset illustrate stretched-exponential fits of the structural decay, Eq.~\eqref{eq-sisf-fit}.
(d) Static and dynamic parameters characterizing the overlap correlation function (CuZr; $T=\SI{850}{\kelvin}$). The normalized static overlap $q_c(z,\infty)/q_c(\infty,\infty)$ (crosses) follows the normalized density profile $\rho(z)/\rho(\infty)$ (line). The normalized change in the relaxation time $\tau_\text{ov}(z)/\tau_\text{ov}(0)$ (squares) is shown in comparison to the corresponding quantity obtained from the $z$-resolved \acrshort{SISF} (circles).
(e) and (f): Position-dependent relative mobility enhancement $\tau(\infty)/\tau(z)-1$ (from the layer-resolved \acrshort{SISF}) for CuZr and the \gls{LJBM}.
}
\label{fg-dynamic}
\end{figure*}

We study two exemplary glass formers by \gls{MD} simulations:
the Kob-Andersen \gls{LJBM} \cite{Kob:1995} showing weak surface layering,
and a model of the molten CuZr alloy with \gls{EAM} many-body interactions \cite{Mendelev:2009} showing strong layering.
Simulations (using the LAMMPS package \cite{lammps})
start in the bulk liquid at high temperature ($T=0.6$ for \gls{LJBM}; $T=\SI{2000}{K}$ for CuZr) and zero pressure. A liquid-vacuum interface was created by an instantaneous increase of the box length along the $z$-axis [see the illustration in Fig.~\ref{fg-dynamic}(a)].
After re-equilibration, the membranes were cooled down to subsequently lower target temperatures in the canonical ensemble (NVT); data
was collected in the microcanonical ensemble (NVE) over 16 realizations per state point.
To check that finite-size effects are irrelevant, we compare simulations of two system sizes: small systems (S) with $L_x=L_y\approx13\sigma$, $L_z\approx31\sigma$ and at least $N=5000$ particles; and large systems (L) with
$L_z\approx40\sigma$ and at least $N=7000$ particles (where $\sigma$ is a typical
atomic size, $\sigma\approx\SI{2.7}{\angstrom}$ for CuZr; precise information is
given as Supplemental Information \cite{si}).
Our simulations are in equilibrium in the sense that all particles are at the same temperature, and no external field is required to maintain the state, once prepared, in the microcanonical ensemble.

The spatially resolved dynamics can be assessed through the overlap correlation
function suggested by the \gls{PTS} method \cite{Cavagna:2007}:
the simulation box is discretized into small cubic units of size $\delta$ (about $0.52\,\sigma\approx\SI{1.4}{\angstrom}$ for CuZr and $0.6\,\sigma$ for the \gls{LJBM}), and the overlap of configurations a time $t$ apart is calculated as:
 $q_c(z,t)={\left\langle \sum_{i}n_i(t)n_i(0)\delta\left(z_i-z\right) \right\rangle}\Big/{\left\langle \sum_{i}n_i(0)\delta\left(z_i-z\right) \right\rangle}$, 
where $n_i=1$ if box $i$ at distance $z_i$ from the surface is occupied by an atom and $n_i=0$ otherwise, and $\langle\cdot\rangle$ denotes an average over the simulation ensemble.

\begin{figure}
\includegraphics[width=0.9\linewidth]{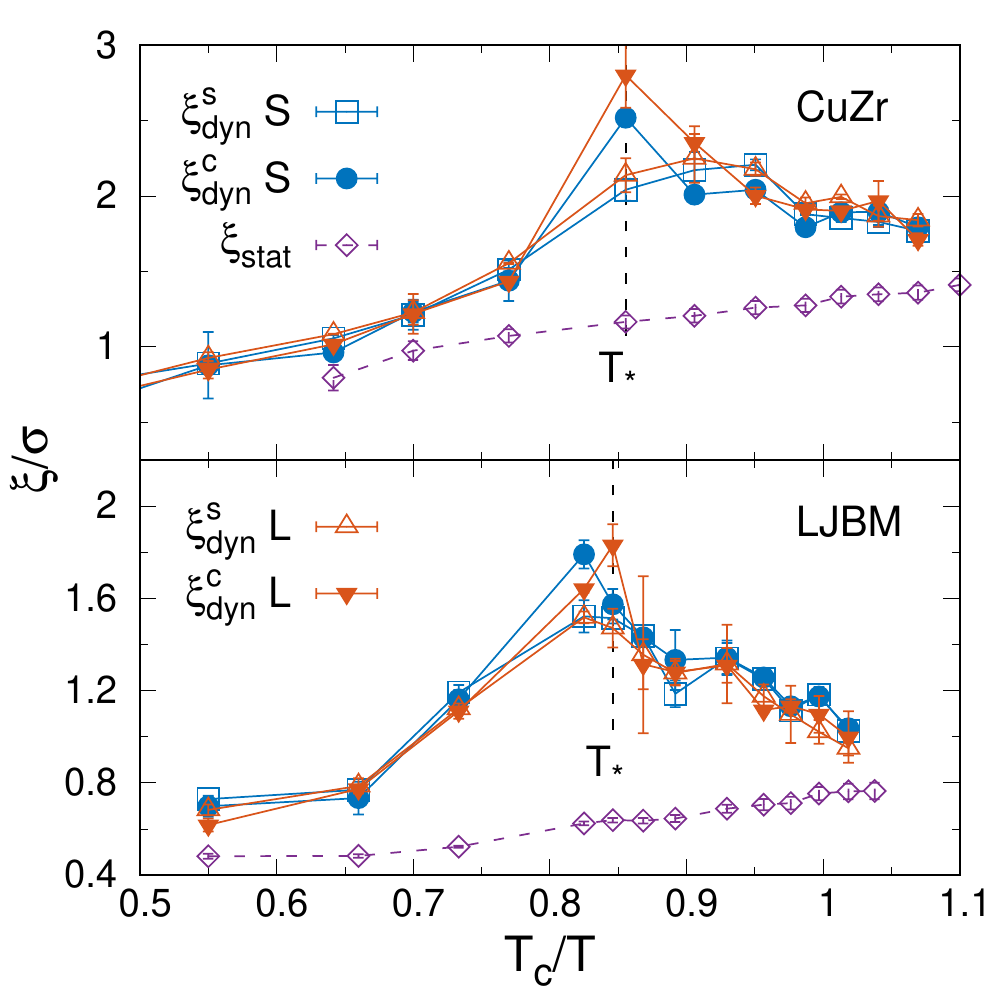}
\caption{Temperature dependent statical $\xi_\text{stat}(T)$ and dynamical correlation lengths $\xi_\text{dyn}(T)$ near a free surface for the CuZr (top panel) and the \gls{LJBM} liquids (bottom panel). The static correlation length $\xi_\textrm{stat}$ is extracted from the exponential decay of $\rho(z)$. Values from the self- ($\xi^\textrm{s}_\textrm{dyn}$) and collective- ($\xi^\textrm{c}_\textrm{dyn}$) intermediate scattering functions are shown in systems of two different sizes (S: small systems; L: large systems). The evolution is non-monotonic around a peak temperature $T_*$ indicated by the dashed vertical lines.
}
\label{fg-tau-length}
\end{figure}

The functions $q_c(z,t)$ follow a standard two-step relaxation pattern 
of dynamical correlation functions near the glass transition [Fig.~\ref{fg-dynamic}(c)]: a short-time relaxation to an intermediate-time plateau is followed by stretched-exponential structural relaxation from the plateau. At long times, $q_c(z,t)$ decays to a non-zero $z$-dependent constant $q_c(z,\infty)$ that represents frozen-in density fluctuations: the introduction of a free surface induces a static density profile $\rho(z)$ [Fig.~\ref{fg-dynamic}(b)], and we find $q_c(z,\infty)\propto\rho(z)$ [Fig.~\ref{fg-dynamic}(d)]. This is the expected behavior for a stationary ergodic system, and it marks an important difference of our system to previous \gls{PTS} analyses where a non-trivial long-time limit of $q_c(z,t)$ signalled a frozen-in nonergodic component of the dynamics.
The static overlap evolves smoothly with decreasing temperature, and
it decays exponentially towards the bulk density; thus, a
\emph{static} correlation length $\xi_\text{stat}$ can be extracted
from fits of the form $q_c(z,\infty)\propto\rho(z)=A(z)\exp(-z/\xi_\text{stat})+\rho(\infty)$, where $\rho(\infty)$ is the density of the bulk liquid.
We use the function $A(z)=A_0\sin(2\pi(z-z_0)/d_p)$ to capture the pronounced
surface-induced layering effects seen for CuZr. They are in agreement with
experiments on metallic \cite{Shpyrko:2004} and nonmetallic liquids \cite{Mo:2006,Haddad:2018}, and grand-canonical \gls{MD} simulations of liquid films \cite{Gao:1997}. The \gls{LJBM} does not show pronounced layering \cite{Chacon:2001},
so that there $A(z)=A_0$ is used.
In both cases, the static length scale $\xi_\text{stat}$ increases monotonically and mildly across the temperature range that we investigate (open symbols in Fig.~\ref{fg-tau-length}). This monotonic increase with decreasing temperature is consistent with the prediction of \gls{RFOT} and with other computer simulation results \cite{Scheidler:2003,Berthier:2011,Flenner:2014}.

To obtain the \emph{dynamical} correlation length,
we parametrize the long-time decay of the overlap correlation function by stretched-exponential functions,
\begin{equation}\label{eq-sisf-fit}
q_c(z,t) = q_0(z)\exp[-(t/\tau_\text{ov}(z))^{\beta(z)}]+q_c(z,\infty)\,, 
\end{equation}
where $\tau_\text{ov}(z)$ is a $z$-dependent relaxation time. Similar fits have been performed for the collective and \gls{SISF}, and we only discuss the features that are robustly displayed by all relaxation times, taking that of the \gls{SISF} as a proxy $\tau(z)$; see Fig.~\ref{fg-dynamic}(d) and Supplemental Material \cite{si}.

In the relative enhancement of the mobility $\mu(z)=1/\tau(z)$, given by $\tau(\infty)/\tau(z)-1$, there emerge two spatial regimes at low temperature [Fig.~\ref{fg-dynamic}(e,f)]:
%
%
%
%
close to the surface ($z\lesssim2\sigma$), an initial exponential decay is identified, whose typical length scale depends only weakly on temperature. This regime corresponds to distances where the static density profile has not yet saturated to its bulk value. An intermediate $z$-range with a much slower decay opens at lower temperatures ($T\lesssim\SI{1000}{\kelvin}$ for CuZr, $T\lesssim0.45$ for LJ). This intermediate regime expands as $T$ is lowered. Here, $\rho(z)\approx\rho(\infty)$, and thus this is the regime where an intrinsic dynamical correlation length $\xi_\text{dyn}$ can be extracted from the exponential decay of $\mu(z)$, viz.
\begin{equation}
\mu(z) = C\exp[-z/\xi_{\text{dyn}}]+\mu(\infty)\,.
\end{equation}
One already anticipates from Fig.~\ref{fg-dynamic}(e--f) that this dynamical correlation length shows a non-monotonic temperature dependence: curves for intermediate temperature (around $T=\SI{810}{\kelvin}$ in the CuZr liquid, and around $T=0.4$ in the LJ binary mixture) extend further into the bulk than those both at higher and at lower temperatures.

The resulting dynamical correlation lengths $\xi_\text{dyn}$ display clear maxima at a temperature $T_*$ (Fig.~\ref{fg-tau-length}).
Both above and below $T_*$, the dynamic and static (symbols with dashed lines in Fig.~\ref{fg-tau-length}) correlation lengths become similar.
In particular, below $T_*$, $\xi_\text{dyn}$ decreases towards the smaller static one, $\xi_\textrm{stat}$, again. This is not a finite-size effect: only around the maximum in $\xi_\text{dyn}$, some slight effects of system size (in line with those expected from conventional four-point correlations in supercooled liquids \cite{Berthier:2012,Peng:2016,Karmakar:2010}) are seen that disappear both at higher and at lower temperatures, and thus give additional evidence that the dynamical correlation length peaks at $T_*$.
In both the CuZr and the \gls{LJBM} system, we note that the peak observed in $\xi_\text{dyn}$ over $\xi_\text{stat}$ is at least a factor of $2$.
It is hence a robust phenomenon across systems with different microscopic interactions and surface features. 


We now demonstrate the intimate link of the maximum in the dynamic correlation length \emph{near the surface} with a cross-over point that governs the \emph{bulk} dynamics. Such a link is remarkable, because the point of maximal correlation length, $T_*$, is clearly above the $T_c$ of \gls{MCT}, to which candidates of structural changes impacting the relevant dynamical regime have so far been linked.
One example is a change in morphology of the \gls{CRR} as suggested by \gls{RFOT} \cite{Stevenson:2006}.

We identify \gls{CRR} as nearest-neighbor clusters of fast particles in simulations of the \emph{bulk} systems. 
Following \cite{Gebremichael:2004,Starr:2013}, fast particles are defined as those that during the time interval corresponding to the average structural relaxation time, have moved significantly farther than what is expected from the average mean-squared displacement.
Clusters are defined by fast particles closer than the first minimum position in the pair distribution function.
To quantify the shape and in particular the anisotropy of these clusters, we consider the ratio of their correlation length to the expected spherical size: In analogy to percolation theory \cite{Nakayama:book}, the average cluster correlation length is given by
\begin{equation}
  \xi_\text{cl}^2=\sum\nolimits_sR_{g,s}^2s^2P(s)/\sum\nolimits_ss^2P(s)\,,
\end{equation}
where the sums run over the individual clusters of size $s$, $P(s)$ is the probability of finding a cluster of size $s$, and $R_{g,s}$ is the radius of gyration of the cluster of size $s$, defined by
$R_{g,s}^2=\frac{1}{2s^2}\left\langle\sum_{ij\in s}(\vec{r}_i-\vec{r}_j)^2\right\rangle_s\,,$
where the sum runs over all particles $i,j$ that are part of the cluster, and $\left\langle \cdots\right\rangle_s$ denotes the average over all clusters of size $s$.
The expected linear dimension of a spherical cluster of size $R_s$ in turn is defined by
$\left\langle s\right\rangle=(4\pi/3)\rho_n R_{s}^3$, where $\rho_n$ is the number density, and $\left\langle s\right\rangle=\sum_{s\ge2}s^2P(s)/\sum_{s\ge2}sP(s)$ is the average cluster size.
The ratio, $\xi_\text{cl}/R_s$, can then be used as a simple proxy to measure the anisotropy of the fast-particle regions.

\begin{figure}
\includegraphics[width=1\linewidth]{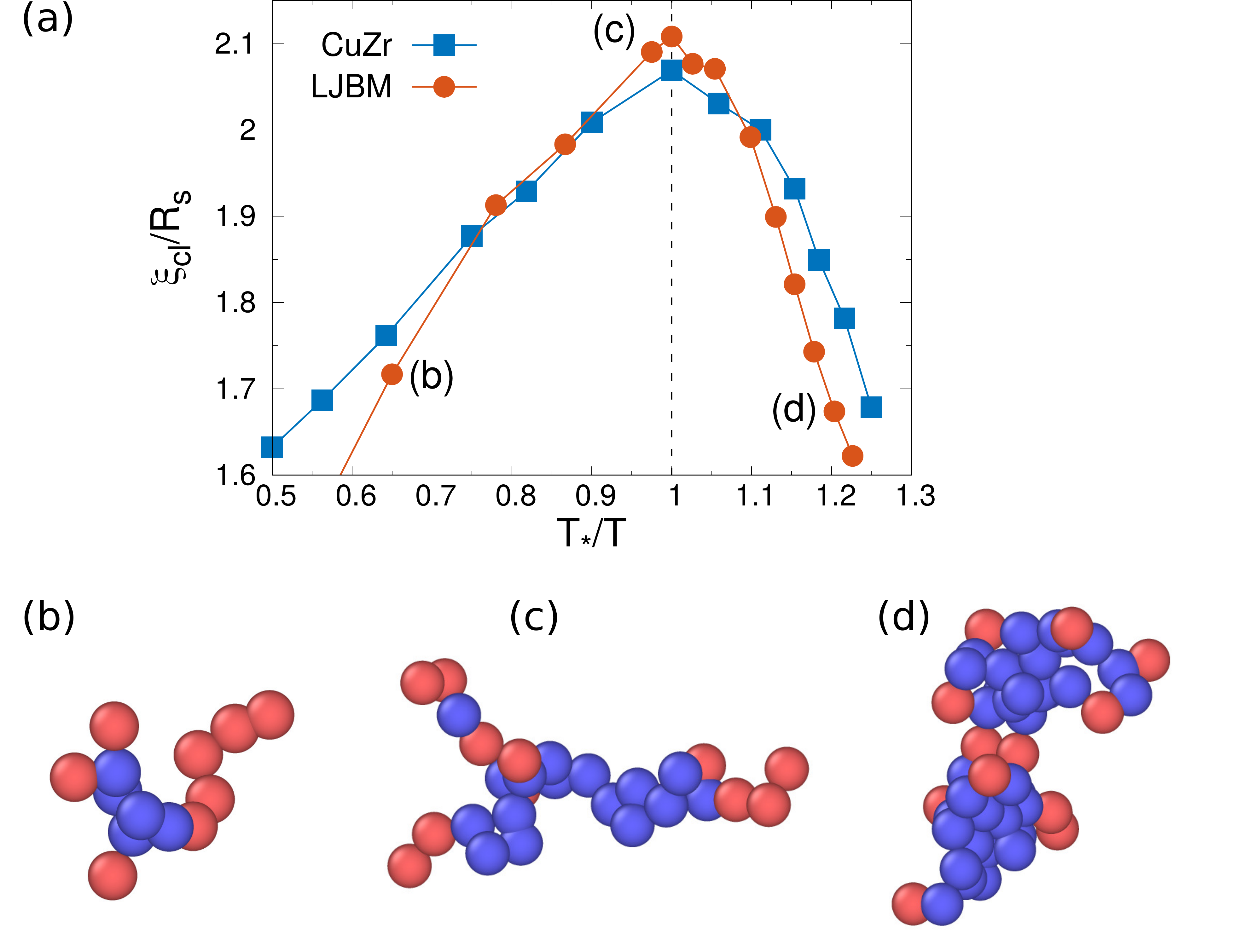}
\caption{Characterization of the \gls{CRR} shape in the bulk liquids.
(a), Aspect ratio of fast-particle clusters in the bulk simulations, $\xi_\text{cl}/R_s$.
The positions of the labels, (b), (c), and (d), correspond to the temperature points at which fast-particle clusters are exemplified in panel (b), (c), and (d), respectively.
Particles in blue correspond to the core of the cluster, defined as having more than two fast nearest neighbours, whereas particles shown in red are those having only one or two fast particles as their nearest neighbors.
}
\label{fg-CRR}
\end{figure}

The aspect ratio of clusters, $\xi_\text{cl}/R_s$, exhibits a striking non-monotonic behaviour with temperature change [Fig.~\ref{fg-CRR}(a)], with a maximum at the same temperature $T_*>T_c$ where also the dynamical correlations near a free surface show a maximum. Thus we argue that $\xi_\text{dyn}$ is intimately related to the shape transition of \gls{CRR} in the bulk.

Typical shapes found for the fast-particle clusters in the bulk are also shown to demonstrate the shape transition [Fig.~\ref{fg-CRR}(b--d)]:
at high temperatures, clusters are small and of a random-walk like fractal structure. As the temperature is lowered, the clusters increase in size, and at temperatures below $T_*$, they are relatively compact objects. Around $T=T_*$, the anisotropy is largest: as the clusters grow in average size upon lowering the temperature, this growth first occurs through a string-like extension of the clusters; only below $T_*$, a more isotropic growth of the clusters is seen. 
While the string-like motion of atoms is well known in supercooled liquids \cite{Donati:1998, Gebremichael:2004,Starr:2013,Betancourt:2014},
this is the first report for the shape of \gls{CRR} transiting back to the compact geometry at low temperatures in atomic glass formers.


\begin{figure}
\includegraphics[width=0.9\linewidth]{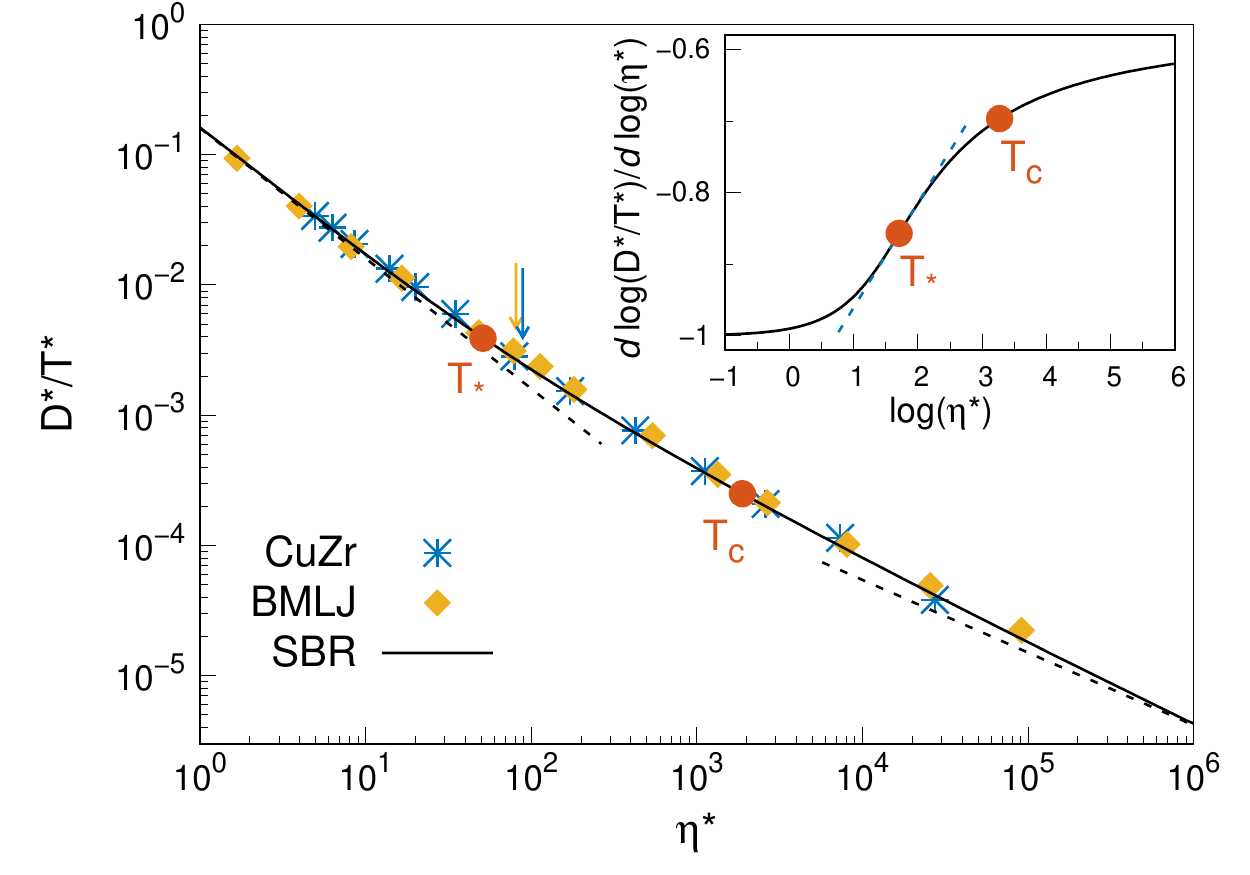}
\caption{Stokes--Einstein plot of diffusivity $D$ versus viscosity $\eta$ (in \gls{LJBM} units) for the \emph{bulk} liquids, compared with the theoretical prediction from \acrfull{SBR} (solid line, using the \gls{MCT} exponent parameter $\lambda=0.75$). 
To make data collapse in different systems, $D/T$ and $\eta$ are scaled to $D^*/T^*$ and $\eta^*$ in CuZr.
Dotted lines are the \gls{SBR} asymptotes for high and low temperatures, i.e., a regular Stokes--Einstein law, $D^*/T^*\sim\eta^{*-1}$, and a fractional law $D^*/T^*\sim\eta^{*-0.558}$. Red circles mark the \gls{MCT}-$T_c$, and $T_*$ predicted from \gls{SBR}; arrows indicate the approximate maximum positions $T_*$ inferred from $\xi_\text{dyn}$ for the two simulated systems.
We determine $T_*$ within \gls{SBR} from the point of maximum slope in the SE crossover curve $d\log(D^*/T^*)/d\log\eta^*$ as shown in the inset.
}
\label{fg-se}
\end{figure}

The emergence of large \gls{CRR} signals heterogeneities in the dynamics that \emph{inter alia} lead to a
breakdown of the \gls{SE} relation \cite{Sengupta:2013,Kawasaki:2013,Glotzer:2000,Peng:2016}: the fast-particle dominated diffusivity decouples from the bulk relaxation that is governed by the slow particles \cite{Sengupta:2014,Schober:2016}. While
far above $T_c$ a \gls{SE} relation for the diffusion coefficient of a tracer particle, $D/T\sim\eta^{-1}$ holds well,
far below $T_c$ a fractional \gls{SE} relation emerges, $D/T\sim\eta^{-x}$ with some exponent $x<1$ (Fig.~\ref{fg-se}).
The maximum anisotropy of \gls{CRR} at $T_*$ suggests that there, the coupling of fast-particle motion to the bulk dynamics changes, and that this can be connected to the breakdown of the \gls{SE} relation \cite{Flenner:2014}.

Crucially, this allows to provide a clear first-principles definition of $T_*$.
We do so by recalling
a recent extension of the asymptotic laws of \gls{MCT}, the \acrfull{SBR} \cite{Rizzo:2013,Rizzo:2014,Rizzo:2016}. It rationalizes the crossover from regular to fractional \gls{SE} relations as arising from long-wave length fluctuations in the local dynamical order parameter \cite{Rizzo:2015}: above $T_c$ the average dynamics is mean-field liquid-like, and the \gls{SE} relation results from the realization $\langle1/\mu\rangle\sim1/\langle\mu\rangle$. Below $T_c$, the relaxation dynamics is dominated by rare fluctuations of liquid regions inside a glass-like matrix, and the tail of the order-parameter distribution gives rise to a fractional \gls{SE} relation, since $\langle1/\mu\rangle\not\sim1/\langle\mu\rangle$. \Gls{SBR} describes our data well (solid line in Fig.~\ref{fg-se}).

Within \gls{SBR}, we can rigorously identify the point of maximal dynamical correlation, $T_*$, as the point where the decoupling of fast-particle dominated motion (diffusivity) from the bulk relaxation is most sensitive to fluctuations in the local glassiness of the dynamics. As \gls{SBR} predicts the logarithmic derivative of $D/T$ as a function of $\eta$ to cross over from $-1$ in the ordinary \gls{SE} regime to $-x$ in the glass, we identify the temperature where this crossover is most rapid as $T_*$ (inset of Fig.~\ref{fg-se}). 
At this point, the competition of dynamic fluctuation between the liquid-like regions and the glass-like ones is the strongest.
This prediction agrees well with the points where the maximum dynamic correlation and the largest anisotropiy of the \gls{CRR} locate (see the marked $T_*$ point and arrows in Fig.~\ref{fg-se}).

In conclusion, we propose supercooled states with a free surface
as a convenient model to interrogate spatial correlations in fully equilibrated systems.
They show a clear separation of the static and the dynamic correlation lengths.
While non-monotonic behavior of a dynamical correlation length was first reported in strongly perturbed systems \cite{Kob:2012,Hocky:2014,Nagamanasa:2015}, we strikingly find the non-monotonicity also emerges in the vapour-liquid interface, a natural system in equilibrium, for the glass formers of pair-wise or many-body interatomic interaction.
The dynamical correlation length displays a clear maximum at a temperature $T_*$ that is associated to the cross-over from liquid-like dynamics to the spatially heterogeneous glass-like dynamics.


We demonstrate that the non-monotonic change in dynamical correlation as measured \emph{near the surface} is linked to a non-monotonic evolution of the aspect ratio of fast-particle clusters \emph{in the bulk}, concomittantly signalled by the breakdown of the Stokes-Einstein relation. Thus the point of maximum dynamical correlations $T_*$ can be identified as the point where the balance of liquid- and glass-like fluctuations in the system is most sensitive to a change in control parameter. In particular the \acrlong{SBR} allows to identify $T_*$ in this way as a temperature that is strongly relevant for the dynamics in glass formers -- both in bulk and for the formation of glassy films -- and that is genuinely distinct from the $T_c$ of \gls{MCT}.

\begin{acknowledgments}
The authors thank Walter Kob and Clemens Bechinger for their valuable comments.
We also appreciate the computer resources provided by 
the High Performance Computing Center (HPC) of Central South University.
\end{acknowledgments}


\bibliographystyle{apsrev4-1}
\bibliography{reference}

\end{document}